\documentstyle[amsfonts,amscd]{amsart}

\numberwithin{equation}{section}

\input amssym.def
\input amssym.def

\begin{document}
\title{Wick type deformation quantization \\
of Fedosov manifolds}
\author{V. A. Dolgushev}
\address[]{Department of Physics, Tomsk State University}
\email[V. A. D]{vald@@phys.tsu.ru}
\author{S. L. Lyakhovich}
\email[S. S. L.]{sll@@phys.tsu.ru}
\author{A. A. Sharapov}
\email[A. A. S.]{sharapov@@phys.tsu.ru}
\keywords{Deformation quantization, Wick symbol, supersymplectic manifolds}
\maketitle

\begin{abstract}
A coordinate-free definition for Wick-type symbols is given for symplectic
manifolds by means of the Fedosov procedure. The main ingredient of this
approach is a bilinear symmetric form defined on the complexified tangent
bundle of the symplectic manifold and subject to some set of algebraic and
differential conditions. It is precisely the structure which describes a
deviation of the Wick-type star-product from the Weyl one in the first order
in the deformation parameter. The geometry of the symplectic manifolds
equipped by such a bilinear form is explored and a certain analogue of the
Newlander-Nirenberg theorem is presented. The $2$-form is explicitly
identified which cohomological class coincides with the Fedosov class of the
Wick-type star-product. For the particular case of K\"ahler manifold this
class is shown to be proportional to the Chern class of a complex manifold.
We also show that the symbol construction admits canonical superextension,
which can be thought of as the Wick-type deformation of the exterior algebra
of differential forms on the base (even) manifold. Possible applications of
the deformed superalgebra to the noncommutative field theory and strings are
discussed.
\end{abstract}




\section{Introduction}

The deformation quantization as it was originally defined in \cite{Ber1},
\cite{BFFLS} has now been well established for every symplectic manifold
through the combined efforts of many authors (for review see \cite{Stern}).
The question of existence of the formal associative deformation for the
commutative algebra of smooth functions, so-called star product, has been
solved by De Wilde and Lecomte \cite{DWL}. The classification of the
equivalence classes of deformation quantization by formal power series in
the second De Rham cohomology has been carried out in several works \cite{NT}%
, \cite{BCG}, \cite{Deligne}, \cite{DWL1}. Finally, in the seminal paper
\cite{Fedosov1} Fedosov has given an explicit geometric construction for the
star product on an arbitrary symplectic manifold making use of the
symplectic connection. As it was shown by Xu \cite{Xu} every deformation
quantization on a symplectic manifold is equivalent to that obtained by
Fedosov's method. Recently, the relationship has been established between
the Fedosov quantization and BRST method for the constrained dynamical
systems \cite{GL}.

In parallel with the general theory of deformation quantization some special
types of star-products, possessing additional algebraic/geometric
properties, have been studied as well. Motivated by the constructions of
geometric quantization and symbol calculus on the K\"ahler manifolds, a
particular emphasis has been placed on the deformation quantization of
symplectic manifolds admitting two transverse polarizations. The respective
quantization constructions may be thought of as a generalization of the Wick
or $qp$-symbol calculus, known for the linear symplectic spaces, much as the
Fedosov deformation quantization may be regarded as a generalization of the
Weyl-Moyal star-product construction. At present, there is a large amount of
literature concerning the deformation quantization on polarized symplectic
manifolds \cite{Mol},\cite{CGR},\cite{Bor1}, \cite{Kar2}, \cite{Kar3}, \cite
{Kar4}, \cite{Mor}, \cite{Pfl}, \cite{RT}, beginning with the pioneering
paper by Berezin \cite{Ber2} on the quantization in complex symmetric spaces.

It should be emphasized that in all the papers cited above the construction
of the Wick-type star-products is based on the explicit use of a special
local coordinate system adapted to the polarization ({\it separation of
variables} in terminology of work \cite{Kar2}). This does not seem entirely
adequate for the physical applications as the most of interesting physical
theories are formulated in general covariant way, i.e. without resorting to
a particular choice of coordinates. So, it is desirable to relate the
polarization with an additional geometric structure (tensor field) on the
symplectic manifold in such a way that the resulting star-product
construction would not imply to use any particular choice of the
coordinates. One may further treat this structure, if necessary, as
dynamical field encoding all the polarizations, just as the Einstein's
equations for metric field govern the geometry of Riemannian manifolds, and
try to assign this structure with a physical interpretation. In this form
the Wick symbols may enjoy some interesting physical applications, two of
which (noncommutative field theories on curved symplectic manifolds and
nonlinear sigma-models) are discussed in the concluding section.

Although it is a commonly held belief that, leaving aside global geometrical
aspects, all the quantizations are (unitary or formally) equivalent to each
other, this equivalence can be spoiled because of quantum divergences which
might appear as soon as one deals with infinite-dimensional symplectic
manifolds (field theories). This is best illustrated by a concrete physical
example of the free bosonic string. Formally, the theory may be quantized by
means of both Weyl and Wick symbols, but the critical dimension of
space-time and nontrivial physical spectrum are known to arise only for the
Wick symbol. This is due to the infinities that appear when one actually
tries to apply the equivalence transformation to the operators of physical
observables. Thus the different quantization schemes may lead to essentially
different quantum mechanics and the Wick symbols are recognized to be more
appropriate for the field-theoretical problems. On the other hand, the
global aspects of quantization cannot be surely ignored in a mathematically
rigor treatment of the question. It is the account of a phase-space topology
that gives a deep geometrical insight into the quantization of such
paramount physical characteristics as the spin, magnetic and electric charge
of the particle, energy levels of hydrogen atom and harmonic oscillator and
so on, both in the frameworks of geometric or deformation quantizations \cite
{Woodhouse}, \cite{FedosovBook}. Finally, even though two quantization are
equivalent in a mathematical sense this does not yet imply their physical
equivalence since, for instance, the same classical observables being
quantized in either case will have different spectra in general \cite{CFGS}.

In this paper we give a constructive coordinate-free definition for the
star-product of Wick type in the framework of the Fedosov deformation
quantization. After the paper \cite{GRS}, a symplectic manifold equipped by
a torsion-free symplectic connection is usually called the Fedosov manifold
because precisely these data, symplectic structure and connection, do enter
the Fedosov star-product \cite{Fedosov1}. The Wick deformation quantization
involves one more geometric structure - a pair of transverse polarizations,
and, by analogy to the previous case, the underlying manifold may be called
as the {\it Fedosov manifolds of Wick type} or {\it FW-manifold }for short.

Let us briefly outline the key idea underlying our approach to the
construction of Wick-type symbols on general symplectic manifolds. Hereafter
the term ``Wick-type'' will be used in a reference to a broad class of
symbols incorporating, along with the ordinary (genuine) Wick symbols, the
so-called $qp$-symbols as well as various mixed possibilities commonly
regarded as the pseudo-Wick symbols. To give a more precise definition of
what is meant here, consider first the linear symplectic manifold ${\Bbb R}%
^{2n}$ equipped with the canonical Poisson brackets $\{y^i,y^j\}=\omega
^{ij} $. Then the usual Weyl-Moyal product of two observables, defined as
\begin{equation*}
a*b(y)=\exp \left( \frac{i\hbar }2\omega ^{ij}\frac \partial {\partial
y^i}\frac \partial {\partial z^j}\right) a(y)b(z)|_{z=y},
\end{equation*}
turns the space of smooth functions in $y$ to the noncommutative associative
algebra with a unit, which is called the algebra of Weyl symbols. Note that
all the coordinates $y$'s enter uniformly in the above formula. Contrary to
this, the construction of Wick-type symbols always implies some (real,
complex or mixed){\it \ polarization} \cite{Woodhouse} splitting the $y$'s
into two sets of (canonically) conjugated variables. For example, the $qp$%
-symbol construction is based on separation of phase-space variables on the
``coordinates'' $q$ and ``momenta'' $p$ (that corresponds to the choice of
some real polarization) and the standard ordering prescription, ``first $q$,
then $p$'', for any polynomial in $y$'s observable. For the complex
polarization the same role is played by pairs of oscillatory variables $q\pm
ip$. Formally, the transition from the Weyl to Wick-type symbols is achieved
by adding a certain complex-valued symmetric tensor $g$ to the Poisson
tensor $\omega $ in the formula for the Weyl-Moyal $*-$product,
\begin{equation}
a*_gb(y)=\exp \left( \frac{i\hbar }2\Lambda ^{ij}\frac \partial {\partial
y^i}\frac \partial {\partial z^j}\right) a(y)b(z)|_{z=y},  \label{wlp}
\end{equation}
\begin{equation*}
\Lambda ^{ij}=\omega ^{ij}+g^{ij},\quad \Lambda ^{\dagger }=-\Lambda ,\quad
i,j=1,2,...,2n.
\end{equation*}
Although the associativity of modified product holds for any constant $g$,
the Wick-type symbols are extracted by the additional condition
\begin{equation}
rank\Lambda =corank\Lambda =n.  \label{rc}
\end{equation}
In particular, the genuine Wick symbol corresponds to a pure imaginary $g$
while the real $\Lambda $ gives a $qp$-symbol. In either case (including the
mixed polarization) the square of the matrix
\begin{equation}
I_j^i=\omega ^{ik}g_{kj}  \label{is}
\end{equation}
is equal to $1$ and hence the operator $I$ generates a polarization
splitting the complexified phase space ${\Bbb C}^{2n}$ in a direct sum of
two transverse subspaces related to the eigen values $\pm 1$.

The formula (\ref{wlp}) may serve as the starting point for the covariant
generalization of the notion of Wick symbol to general symplectic manifolds.
Turning to the curved manifold $M$, $\dim M=2n$, we just replace the
constant matrix $\Lambda $ by a general complex-valued bilinear form $%
\Lambda (x)=\omega (x)+g(x)$ with the closed and non-degenerate
antisymmetric part $\omega (x)$ and satisfying to the above half-rank
condition (\ref{rc}). Then each tangent space $T_xM,\quad x\in M,$ turns to
the symplectic vector space w.r.t. $\omega (x)$ and may be quantized by
means of Wick-type product (\ref{wlp}). Taking the union of all tangent
spaces equipped with the star-product we get the bundle of Wick symbols,
which is a sort of ``quantum tangent bundle''. Then, following Fedosov's
idea, we introduce a flat connection on it by adding some quantum correction
to the usual affine connection preserving $\Lambda (x)$. The flat sections
of this connection can be naturally identified with the space of quantum
observables $C^\infty (M)[[\hbar ]]$. Finally, the pull back of the bundle
star-product via the Fedosov connection induces a star-product on $C^\infty
(M)[[\hbar ]]$. The only crucial point of this program is the existence of a
torsion-free linear connection $\nabla $ preserving $\Lambda $. As we show
bellow the necessary and sufficient condition for such a connection to exist
is the integrability of the right and left kernel distribution of $\Lambda
(x).$ When the latter condition is fulfilled $\nabla $ is just the
Levi-Civita connection associated to the symmetric and non-degenerate form $%
g(x)$ and the right and left kernel distributions define the transverse
polarization of the symplectic manifold $(M,\omega )$. Under specified
conditions we refer to the pair $(M,\Lambda )$ as the FW-manifold.

The paper is organized as follows. In Section 2 we define the Fedosov
manifolds of Wick type, discuss their geometry and give some examples. The
main tool we use here is the integrable involution structure (\ref{is})
associated to any FW-manifold structure $\Lambda $. The deformation
quantization on the FW-manifolds by Fedosov's method is presented in Section
3. In Section 4 we pose the question about equivalence between original
Fedosov star-product (generalized Weyl symbols) and star-product of
Wick-type. The answer is follows: The only obstruction for the equivalence
is associated with a non-zero De Rham class of a certain (in general
complex) 2-form, which explicit expression is given by contraction of the
curvature tensor of the FW-manifold and the corresponding involution tensor.
In the K\"ahler case, this 2-form represents the first Chern class $c_1$ of
a complex manifold. Section 5 is devoted to the superextension of the
previous constructions to supersymplectic manifold, which, as we show, can
be canonically associated to the tangent bundle of any FW-manifold. This
superextension is a particular example of more general construction of
super-Poisson brackets and their quantization proposed some time ago by\
Bordemann \cite{SBor}. Here we also study the relationship between the
algebra of quantum observables on initial FW-manifold and that on its
superextension. This relationship is not so evident because the canonical
projection of superextended FW-manifold to the base $(M,\Lambda )$ or the
natural embedding of the FW-manifold to its superextension do not induce
homomorphism of the corresponding algebras.

\section{Fedosov manifolds of Wick type.}

Consider a $2n$-dimensional real manifold $M$ equipped with a complex-valued
bilinear form $\Lambda $ (not necessarily symmetric or antisymmetric). In
local coordinates $\{x^i\}$ on $M$ the form $\Lambda $ is completely
determined by its components $\Lambda _{ij}=\Lambda (\partial _i,\partial
_j) $ , where $\partial _i\equiv \partial /\partial x^i.$ Having the form $%
\Lambda $ one can define two maps from the comlexified tangent bundle to the
comlexified cotangent one: for any $X\in TM$ the corresponding linear forms
are 
\begin{equation}
\Lambda (\cdot ,X)\ ,\qquad \Lambda ^t(\cdot ,X)=\Lambda (X,\cdot )
\label{a1}
\end{equation}
$\Lambda ^t$ being transpose to $\Lambda $. Denote by $\ker \Lambda $ and $%
\ker \Lambda ^t$ the right and left kernel distributions of $\Lambda $.
Obviously, $\dim \ker \Lambda =$ $\dim \ker \Lambda ^t$ . It is convenient
to decompose $\Lambda $ into the sum of symmetric and antisymmetric parts 
\begin{equation}
\Lambda =\omega +g~,\quad \omega =\frac 12(\Lambda -\Lambda ^t)~,\quad
g=\frac 12(\Lambda +\Lambda ^t)  \label{a2}
\end{equation}

\vspace{0.5cm} \noindent
{\bf Definition 1.1} {\it The pair }$(M,\Lambda )${\it \ will be called by
an almost Fedosov manifold of Wick type (or almost FW-manifold for short) if
at each point }$p\in M$

{\it i) }$\omega =\frac 12(\Lambda -\Lambda ^t)${\it \ is a }real{\it \
non-degenerate two-form,}

{\it ii) }$\dim _{{\Bbb {C}}}\ker \Lambda =\frac 12\dim M${\it .}

\vspace{0.5cm}\noindent
The first condition merely implies that the antisymmetric part of $\Lambda $
defines on $M$ an {\it almost symplectic structure }$\omega $ on $M$. From
the second condition it then follows, that at each point $x\in M,$ the left
and right kernel distributions span the tangent space $T_x^{{\Bbb {C}}}M$ .
Indeed, the conditions the vector fields $X~(Y)~$belong to the left (right)
kernel distributions, $\Lambda (X,\cdot )=0~,\quad \Lambda (\cdot ,Y)=0,$
can be rewritten in the form: 
\begin{equation}
IX=X~,\qquad IY=-Y,  \label{a3}
\end{equation}
where the smooth field of automorphisms $I(x):T_x^{{\Bbb C}}M\rightarrow
T_x^{{\Bbb C}}M$ is defined, in any coordinate chart on $M$, by the matrix

\begin{equation*}
I_j^i=\omega ^{ik}g_{kj},
\end{equation*}
with $\omega ^{ij}$ being inverse to $\omega _{ij}$, $\omega ^{ik}\omega
_{kj}=\delta _j^i$. Belonging to the different eigenvalues ($\pm 1$) of the
map $I$, the vectors $X$ and $Y$ are linearly independent and hence $\ker
\Lambda \cap \ker \Lambda ^t=0.$ Accounting the dimensions of the kernel
distributions one concludes that $\ker \Lambda (x)\oplus \ker \Lambda
^t(x)=T_x^{{\Bbb C}}M$, $\forall x\in M.$ Additionally, we have proved that
the form $g$ is non-degenerate and the automorphism $I$ is {\it involutive},
i.e. 
\begin{equation}
I^2=id~(identical~transformation)
\end{equation}
That is why we will call $I$ by an {\it almost involution structure.} Note
that for an anti-Hermitian $\Lambda $, that is $\Lambda ^t=-\overline{%
\Lambda }$, the components of tensor $I$ are purely imaginary 
\begin{equation}
I=\sqrt{-1}J\quad ,\qquad J^2=-id
\end{equation}
and hence they define (and defined by) an almost complex structure $J$. As
is well known the almost complex structure becomes the complex one if it
defines a structure of the complex manifold. Due to the Newlander-Nirenberg
identification theorem for the complex manifolds \cite{NN} it is equivalent
to vanishing of the Nijenhuis tensor associated to $J$ providing the latter
is sufficiently smooth. The vanishing of the Nijenhuis tensor is known to be
equivalent, in turn, to the integrability of vector distributions belonging
to the eigenvalues $\pm \sqrt{-1}$ of $J$. As we will see bellow the analog
of the latter condition can be advocated for the almost involution structure 
$I$ as well, with precisely the same definition of the Nijenhuis tensor.

\vspace{0.5cm} \noindent
{\bf Definition 1.2 }{\it The Nijenhuis tensor associated to the almost
involution structure }$I${\it \ is a smooth tensor field of skew-symmetric
bilinear maps }$N:T_x^{{\Bbb C}}M\wedge T_x^{{\Bbb C}}M\rightarrow T_x^{%
{\Bbb C}}M${\it , which can be defined in two equivalent ways:}

{\it i) for every pair of smooth vector fields }$X${\it \ and }$Y${\it \ } 
\begin{equation}  \label{n1}
N(X,Y)=[X,Y]-I[IX,Y]-I[X,IY]+[IX,IY]~,
\end{equation}

{\it where the brackets }$[\cdot ,\cdot ]${\it \ stand for the commutator of
vector fields;}

{\it ii) let }$\nabla ${\it \ be an arbitrary torsion-free connection, then
in local coordinates }$\{x^i\}$

{\it the components of }$N${\it \ are given by } 
\begin{equation}
N_{ij}^k=I_i^l\nabla _lI_j^k-I_j^l\nabla _lI_i^k-I_l^k(\nabla _iI_j^l-\nabla
_jI_i^l)  \label{n2}
\end{equation}
One may easily check that the relations (\ref{n1},\ref{n2}) do really define
(the same) tensor as they do not actually depend on derivatives of $X,Y$ and
on the choice of connection $\nabla .$ Before examining the geometric
consequences of the condition $N=0,$ let us introduce the key ingredient of
our construction.

\vspace{0.5cm} \noindent
{\bf Definition 1.3 }{\it The almost FW-manifold }$(M,\Lambda )${\it \ will
be called the FW-manifold if there exist a torsion-free connection }$\nabla $%
{\it \ preserving }$\Lambda ${\it \ , i.e. }$\nabla \Lambda =0.$

\vspace{0.5cm} \noindent
Obviously, the form $\Lambda $ is covariantly constant iff both its
symmetric and antisymmetric part is the constant, i.e. 
\begin{equation*}
\nabla \Lambda =0\Leftrightarrow \nabla \omega =\nabla g=0.
\end{equation*}
As there is the only torsion-free connection compatible with a given
non-degenerate symmetric form $g$, the existence of such a connection $%
\nabla $ implies its uniqueness. On the other hand, the totally
antisymmetric part of the equation $\nabla \omega =0$ written in some local
coordinates suggests that the form $\omega $ is closed, i.e. $d\omega =0,$
and therefore any FW-manifold is a symplectic manifold as well. Emphasize
that in general we deal with the complex-valued form $g$, so the connection $%
\nabla $, when exists, is supposed to act in the comlexified tangent bundle
and determined, in each coordinate chart, by complex-valued Christoffel
symbols. The following theorem gives the explicit criteria for an almost
FW-manifold $(M,\Lambda )$ to admit a torsion-free connection preserving $%
\Lambda $.

\vspace{0.5cm} \noindent
{\bf Theorem 1.1 }{\it Given an almost FW-manifold }$(M,\Lambda )${\it \
with the closed antisymmetric part of }$\Lambda ${\it , then the following
statements are equivalent:}

{\it i) }$\Lambda ${\it \ defines the structure of FW-manifold,}

{\it ii) the involution }$I${\it \ associated to }$\Lambda ${\it \ has the
vanishing Nijenhuis tensor,}

{\it iii) the kernel distributions }$\ker \Lambda ${\it \ and }$\ker \Lambda
^t${\it \ are integrable.}

\noindent
Although these assertions are quite elementary we prove them below as they
have a direct bearing on the deformation quantization of FW-manifolds, which
will be considered in the next sections.

\vspace{0.5cm} \noindent
{\bf Proof.} We will proceed following the scheme: $i)\Leftrightarrow
ii),\quad ii)\Leftrightarrow iii).$

\noindent
Implication $i)\Rightarrow ii)$ straightforwardly follows from the second
definition of Nijenhuis tensor (\ref{n2}) in which $\nabla $ is taken to be
compatible with $\Lambda $ (and hence with $I$ ). Conversely, let $\nabla $
be the torsion-free connection preserving $g$. Accounting that $d\omega =0,$
the expression (\ref{n2}) for the Nijenhuis tensor can be reduced to 
\begin{equation*}
N_{jk}^i=2\omega ^{il}\nabla _l\omega _{jk},
\end{equation*}
from which the implication $ii)\Rightarrow i)$ is immediately follows. Now
let $X$ and $Y$ be two eigenvector fields of the involution tensor with the
same eigenvalue $\alpha $; of course, $\alpha ^2=1.$ Evaluating $N$ on these
vector fields with the help of first definition (\ref{n1}) we get 
\begin{equation}
N(X,Y)=2([X,Y]-\alpha I[X,Y])  \label{n3}
\end{equation}
So, if $N=0$ then $I[X,Y]=\alpha [X,Y]$ and the eigen distributions of $I$
are involutive. This proves the implication $iii)\Rightarrow ii)$. The
relation (\ref{n3}) also implies that $N$ comes to zero on each pair of
tangent vectors belonging to the same kernel distribution, $\ker \Lambda $
or $\ker \Lambda ^t$, whenever they are integrable. In the case of $X$ and $%
Y $ belonging to the different distributions associated to the eigenvalues $%
\alpha $ and -$\alpha $, respectively, the value of the Nijenhuis tensor on
them is equal to zero identically: 
\begin{eqnarray*}
N(X,Y) &=&[X,Y]-I[IX,Y]-I[X,IY]+[IX,IY]~= \\
&& \\
&=&[X,Y]-\alpha I[X,Y]+\alpha I[X,IY]+[X,IY]=0~
\end{eqnarray*}
Since the vectors from left and right kernel distributions form the basis of 
$T_xM,$ $\forall $ $x\in M$, the last conclusion proves the implication $%
ii)\Rightarrow iii)$ and the theorem.

\vspace{0.5cm} \noindent
{\bf Examples.} The extended list of examples is provided by the K\"ahler
manifolds. In this case the form $\Lambda $ is anti-Hermitian and the
integrable involution structure $I$ is identified with the complex one
multiplied by $\sqrt{-1}$. In the frame $\{\partial /\partial z^a,\partial
/\partial \overline{z}^b\}$ associated to the local holomorphic coordinates $%
\{z^a\},a=1,...,n,$ on $M$ the matrix of the form $\Lambda $ looks like 
\begin{equation*}
\Lambda =\left( 
\begin{array}{cc}
0 & h_{a\overline{b}} \\ 
0 & 0
\end{array}
\right)
\end{equation*}
where $h=h_{a\overline{b}}dz^a\wedge d\overline{z}^b$ is the K\"ahler
(1,1)-form, $dh=0.$ The form $\Lambda $ is preserved by the K\"ahler
connection with non-zero components of the Christoffel symbols being $\Gamma
_{bc}^a=h^{a\overline{d}}\partial h_{b\overline{d}}/\partial z^c=\overline{%
(\Gamma _{\overline{b}\overline{c}}^{\overline{a}})}$, the matrix $h^{%
\overline{d}a}$ is inverse to $h_{a\overline{b}}$. In particular, all
two-dimensional orientable manifolds admit the structure of FW-manifold as
they known to admit the K\"ahler structure. More generally, let $g$ be a
(pseudo-)Riemannian metric on a two-dimensional real manifold and let $%
\omega =\sqrt{|\det g|}dx^1\wedge dx^2$ be the corresponding volume form.
Then the form 
\begin{equation*}
\Lambda =g+\alpha \omega
\end{equation*}
is obviously preserved by the metric connection for any $\alpha \in {\Bbb {C}%
}$. In order for $\Lambda $ to define the structure of FW-manifold we have
to require 
\begin{equation*}
\det \Lambda =\det g+\alpha ^2|\det g|=0,
\end{equation*}
that fixes $\alpha =\pm \sqrt{-1}$ for the Riemannian metric ($\det g>0)$
and $\alpha =\pm 1$ for the pseudo-Riemannian one ($\det g<0$). The former
case corresponds to the K\"ahler manifolds, while the latter is concerned
with the manifolds endowed by a real structure. The last situation is
strikingly illustrated by the example of one-sheet hyperboloid embedded into
three dimensional Minkowsky space as the surface 
\begin{equation*}
x^2+y^2-z^2=1
\end{equation*}
$(x,y,z)$ being linear coordinates in $R^{2,1}$. The induced metric is of
the pseudo-Euclidean type and it has a constant negative curvature. For the
corresponding real structure, the integral leaves of the eigen distributions
coincide with the two transverse sets of linear elements of the hyperboloid.
Obviously, these linear elements are nothing but the isotropic geodesics of
metric $g$.

Let us recall that the integrable complex distribution $P$ on a symplectic
manifold $(M,\omega )$ is called {\it polarization} if $\dim _{{\Bbb {C}}%
}P=\dim M/2$ and $\omega _{|_P}=0$. In other words, at each point $x\in M$, $%
P$ extracts a (complex) Lagrangian subspace $P_x\subset T_x^{{\Bbb {C}}}M$
in the complexified tangent bundle. It is easy to see that the right and
left kernel distributions on the FW-manifold $(M,\Lambda )$ are Lagrangian
and hence define a pair of transversal polarizations $P_R,$ $P_L$. These
polarizations are highly important for physical applications both in the
framework of the deformation quantization and geometric one as it allows one
to introduce the notion of {\it a} {\it state} of quantum-mechanical system.
The quantization on symplectic manifolds endowed with a pair of transversal
polarizations was intensively studied in two limiting cases: $P_R=\overline{P%
}_L,~P_R\cap $ $\overline{P}_L=0.$ The first possibility is realized for the
K\"ahler manifolds with holomorphic-anti-holomorphic polarizations, while
the second implies the existence of a pair of transversal Lagrangian
foliations on $M$, as a consequence of the Frobenius theorem. It is
interesting to note that for the real $\Lambda =g+\omega $ the leaves of
foliations $P_R,$ $P_L$, being Lagrangian submanifolds with respect to the
symplectic structure $\omega ,$ turn out to be {\it totally geodesic
submanifolds} with respect to pseudo-Riemannian structure $g$. The proof of
the last fact is straightforward and we omit it.

\section{Deformation quantization on FW-manifolds.}

Let, as before, $(M,\Lambda )$ be an FW-manifold of dimension $2n$. We are
reminded that $\Lambda $ can be splitted into the sum of symmetric and
antisymmetric parts (\ref{a2}) which are both non-degenerate. Introduce the
second-rank contravariant tensor field $\Lambda ^{ij}(x)\partial _i\otimes
\partial _j$ defined as follows: 
\begin{equation*}
\Lambda ^{ij}=\omega ^{im}\Lambda _{mn}\omega ^{nj}=g^{ij}+\omega ^{ij},
\end{equation*}
where $\omega ^{ij}$ and $g^{ij}$ are matrices inverse to $\omega _{ij}$ and 
$g_{ij}$, respectively. By construction, 
\begin{equation*}
rank(\Lambda ^{ij})=rank(\Lambda _{ij})=n
\end{equation*}
and $\omega ^{ij}$ is a Poisson tensor. Under the deformation quantization
of the FW-manifold $(M,\Lambda )$ we will mean the construction of an
associative multiplication operation $*$ of two functions, which is an
one-parametric deformation of the ordinary pointwise multiplication in the
algebra $C^\infty (M)$ and which meets the ``boundary condition'': 
\begin{equation}
a*b=ab-\frac{i\hbar }2\Lambda ^{ij}\partial _ia\partial _jb+\ldots
\label{bb1}
\end{equation}
where $\hbar $is the formal deformation parameter (``Plank constant''), and
dots mean the terms of higher orders in $\hbar .$ The condition (\ref{bb1})
is compatible with so-called {\it correspondence principle }of quantum
mechanics: 
\begin{equation}
\lim_{\hbar \rightarrow 0}\frac i\hbar (a*b-b*a)=\{a,b\},  \label{bb2}
\end{equation}
where $\{\cdot ,\cdot \}$ means the Poisson brackets associated to $\omega
^{ij}$. The boundary condition should also be added by the requirement of
locality 
\begin{equation}
supp(f*g)=supp(f)\cap supp(g)  \label{bb3}
\end{equation}
To meet the latter condition the coefficient of each power of $\hbar $ in (%
\ref{bb1}) are usually restricted to be a finite-order differential
expression bilinear in $a$ and $b.$ Note that when $\hbar $ is treated as a
formal (not numerical) parameter the $*$-product of two functions is not a
function but is an element of a more wide space $C^\infty (M)[[\hbar ]]$
consisting of the formal series: $a(x,\hbar )=a_0(x)+\hbar a_1(x)+\hbar
^2a_2(x)+\ldots ,~a_i(x)\in C^\infty (M)$. The space $C^\infty (M)[[\hbar ]]$
is closed already with respect to $*$-multiplication and it may be regarded
as the {\it algebra of} {\it quantum observables, }much as the Poisson
algebra of smooth functions $C^\infty (M)$ on symplectic manifold $(M,\omega
)$ is identified with the space of classical observables. The problem now is
to construct an associative $*$-product starting from the ansatz (\ref{bb1}%
). In this section we show how this the problem can be resolved by a minimal
modification of Fedosov's geometric approach to the deformation quantization 
\cite{Fedosov1}.

\vspace{0.5cm}\noindent
{\bf Definition 2.1} {\it The formal symbol algebra }${\cal A}=\oplus
_{m,n=0}^\infty {\cal A}_{m,n}${\it \ is the bi-graded associative algebra
over }${\Bbb {C}}${\it \ with a unit whose elements are formal series} 
\begin{equation}
a(x,y,dx,\hbar )=\sum_{2k+p,q\ge 0}\hbar ^ka_{k\,,i_1\ldots i_pj_1.\ldots
j_q}(x)y^{i_1}\ldots y^{i_p}dx^{j_1}\wedge \ldots \wedge dx^{j_q}  \label{b1}
\end{equation}
{\it Here the expansion coefficients }$a_{k,\,i_1\ldots i_pj_1\ldots j_q}(x)$%
{\it \ are the components of covariant tensor fields on }$M${\it \ symmetric
with respect to }$i_1,\ldots ,i_p${\it \ and antisymmetric in }$j_1,\ldots
,j_q${\it , }$\hbar ${\it \ is a formal (deformation) parameter; and }$%
\{y^i\}${\it \ are variables transforming as components of tangent vector,
hence the whole expression (\ref{b1}) does not depend on the choice of
coordinates. The general term of the expansion (\ref{b1}) is assigned by a
bi-degree }$(2k+p,q)${\it \ and thus belongs to the subspace }${\cal A}%
_{2k+p,q}${\it . The product of two elements }$a,b\in {\cal A}${\it \ is
defined by the rule} 
\begin{equation}
a\circ b=exp\left( \frac{i\hbar }2\Lambda ^{ij}(x)\frac \partial {\partial
y^i}\frac \partial {\partial z^j}\right) a(x,y,dx,\hbar )\wedge
b(x,z,dx,\hbar )|_{z=y}~,  \label{b2}
\end{equation}
{\it where }$\wedge ${\it \ stands for the ordinary exterior product of
differential forms.}

\vspace{0.5cm}\noindent
It is easily seen that the multiplication (\ref{b2}) is associative and
bi-graded, i.e. ${\cal A}_{m,n}\circ {\cal A}_{k,l}\subset {\cal A}%
_{m+k,n+l} $. In what follows we will refer to subscripts $m$ and $n$
labelling the graded subspace ${\cal A}_{m,n}$ as the first and second
degree respectively. The natural filtration in ${\cal A}$ with respect to
the first degree, 
\begin{equation}
{\cal A}\supset {\cal A}_{\bullet }^1\supset {\cal A}_{\bullet }^2\supset
\ldots ,\quad {\cal A}_n^m\equiv \bigoplus\limits_{k\geq m}{\cal A}_{k,n},
\label{b3}
\end{equation}
defines the topology and convergence in the space of infinite formal series (%
\ref{b1}). Being associative algebra, ${\cal A}$ can be turned to a
differential graded Lie algebra with respect to the second degree: the
commutator of two homogeneous elements $a\in {\cal A}_{m,n},b\in {\cal A}%
_{k,l}$ is given by $[a,b]=a\circ b-(-1)^{nl}b\circ a,$ and extends to whole 
${\cal A}$ by linearity; the nilpotent differential $\delta :{\cal A}%
_{m,n}\rightarrow {\cal A}_{m-1,n+1}$ acts as 
\begin{equation}
\delta a=dx^k\wedge \frac{\partial a}{\partial y^k},\quad \delta (\delta
a)=0,\quad \forall a\in {\cal A}  \label{b4}
\end{equation}
Alternatively, one can write 
\begin{equation}
\delta a=-\frac 1{i\hbar }[\omega _{ij}y^idx^j,a],  \label{b6}
\end{equation}
and hence $\delta $ is an inner derivation of the superalgebra and of the $%
\circ $-product as well: 
\begin{equation}
\delta (a\circ b)=(\delta a)\circ b+(-1)^na\circ (\delta b),\quad \forall
a\in {\cal A}_{\bullet ,n},~\forall b\in {\cal A}  \label{b7}
\end{equation}
Note that $\delta $ acts ``algebraically'' in a sense that it does not
involve derivatives with respect to $x$. The nontrivial cohomologies of $%
\delta $ corresponds to the subspace of quantum observables $C^\infty
(M)[[\hbar ]]\subset {\cal A}$ whose elements are independent of $y$ and $dx$%
; for the complementary subspace one can construct a homotopy operator $%
\delta ^{-1}:{\cal A}_{m,n}\rightarrow {\cal A}_{m+1,n-1}$ of the form 
\begin{equation}
\delta ^{-1}a=y^ki\left( \frac \partial {\partial x^k}\right)
\int\limits_0^1a(x,ty,tdx,\hbar )\frac{dt}t,  \label{b9}
\end{equation}
where $i(\partial /\partial x^k)$ means the contraction of the vector field $%
\partial /\partial x^k$ and a form. Extending the action of $\delta ^{-1}$
to $C^\infty (M)[[\hbar ]]$ by zero, we get the ``Hodge-De Rham
decomposition'' holding for any $a\in {\cal A}$, 
\begin{equation}
a=\sigma (a)+\delta \delta ^{-1}a+\delta ^{-1}\delta a\,,  \label{b10}
\end{equation}
where $\sigma (a)=a(x,0,0,\hbar )$ denotes the projection of $a$ onto $%
C^\infty (M)[[\hbar ]]$.

\vspace{0.5cm}\noindent
Given the torsion-free connection $\nabla $ preserving FW-manifold structure 
$\Lambda $, it induces the covariant derivative on elements of ${\cal A}$
which will be denoted by the same symbol, 
\begin{equation}  \label{b11}
\nabla :{\cal A}_{n,m}\rightarrow {\cal A}_{n,m+1},\quad \nabla =dx^i\wedge
\left( \frac \partial {\partial x^i}-y^j\Gamma _{ij}^k(x)\frac \partial
{\partial y^k}\right) ,
\end{equation}
$\Gamma _{ij}^k$ are Christoffel symbols of the connection $\nabla $.
Definition 1.3 implies the following property of the covariant derivative: 
\begin{equation}  \label{b12}
\nabla (a\circ b)=\nabla a\circ b+(-1)^na\circ \nabla b,\quad \forall a\in 
{\cal A}_{\bullet ,n},\forall ~b\in {\cal A}
\end{equation}
The next Lemma is a counterpart of Lemma 2.4 in \cite{Fedosov1} for
FW-manifolds.

\vspace{0.5cm}\noindent
{\bf Lemma 2.1.}{\it \ Let }$\nabla ${\it \ be the covariant derivative of }$%
{\cal A}${\it . Then} 
\begin{equation}
\nabla \delta a+\delta \nabla a=0,
\end{equation}
\begin{equation}
\nabla ^2a=\nabla (\nabla a)=\frac 1{i\hbar }\,[\,R\,,\,a\,],\qquad R=\frac
14R_{ijkl}y^iy^jdx^k\wedge dx^l,
\end{equation}
{\it where }$R_{ijkl}=\omega _{im}R_{~jkl}^m${\it \ is the curvature tensor
of the connection }$\nabla ${\it .}

\vspace{0.5cm}\noindent
{\bf Proof. }The first identity holds because $\nabla $ is a symmetric
connection. It follows from the definition (\ref{b11}) that 
\begin{equation}
\nabla ^2a=\frac 12dx^k\wedge dx^lR_{ikl}^jy^i\frac{\partial a}{\partial y^j}
\end{equation}
So, $\nabla ^2$ is an algebraic operator. On the other hand, 
\begin{eqnarray*}
\frac 1{i\hbar }[\,R\,,a] &=&-\frac 14(R_{ijkl}y^i\Lambda ^{jn}\frac{%
\partial a}{\partial y^n}-R_{ijkl}y^i\Lambda ^{nj}\frac{\partial a}{\partial
y^n})dx^k\wedge dx^l+ \\
&& \\
&&+\frac{i\hbar }8(R_{ijkl}\Lambda ^{jm}\Lambda ^{in}\frac{\partial ^2a}{%
\partial y^m\partial y^n}-R_{ijkl}\Lambda ^{mj}\Lambda ^{ni}\frac{\partial
^2a}{\partial y^m\partial y^n})dx^k\wedge dx^l \\
&& \\
&=&-\frac 12dx^k\wedge dx^lR_{jikl}y^i\omega ^{nj}\frac{\partial a}{\partial
y^n}=\nabla ^2a
\end{eqnarray*}
Here we have used the Ricci identity, $~g_{in}R_{jkl}^n=-g_{jn}R_{ikl}^n,$
and its symplectic analog $\omega _{im}R_{~jkl}^m=\omega _{jm}R_{~ikl}^m$
(for the proof see (\cite{GRS}). The Lemma proved suggests that the square
of external derivative $\nabla $ is a derivative again and, moreover, it is
an inner derivative of the algebra ${\cal A}$. This fact is of primary
importance for the $*$-product construction along the line of the Fedosov
approach.

\vspace{0.5cm}\noindent
Following Fedosov, we define a more general derivative in ${\cal A}$ of the
form 
\begin{equation}
D=\nabla -\delta +\frac 1{i\hbar }[r,\cdot ]=\nabla +\frac 1{i\hbar }[\omega
_{ij}y^idx^j+r,\cdot ],  \label{b13}
\end{equation}
where $r=r_i(x,y,dx,\hbar )dx^i$ belongs to ${\cal A}_1^3$ and satisfies the
Weyl normalization condition\\$r_i(x,0,dx,\hbar )dx^i=0$. A simple
calculation yields that 
\begin{equation}
D^2a=\frac 1{i\hbar }[\Omega ,a],\quad \forall a\in {\cal A},  \label{b14}
\end{equation}
where 
\begin{equation}
\Omega =-\frac 12\omega _{ij}dx^i\wedge dx^j+R-\delta r+\nabla r+\frac
1{i\hbar }r\circ r  \label{b15}
\end{equation}
is the curvature of $D$. A connection of the form (\ref{b13}) is called {\it %
Abelian }if two-form $\Omega $ does not contain $y$'s, i.e. belongs to the
center of ${\cal A}$. It is clear, that the kernel subspace of an Abelian
connection $D$ is automatically a subalgebra in ${\cal A}$. Denote ${\cal A}%
_D=\ker D\cap {\cal A}_{\bullet ,0}$. The following two theorems are pivot
for the $*$- product construction and they can be proved completely
analogously to Fedosov's original theorems in \cite[Theorems 3.2, 3.3]
{Fedosov1}.

\vspace{0.5cm}\noindent
{\bf Theorem 2.1. }{\it There exists a unique }$r\in ${\it \ }${\cal A}_1^3$%
{\it \ obeying condition }$\delta ^{-1}r=0${\it \ such that }$D${\it , given
by (\ref{b13}), is an Abelian connection with the curvature }$\Omega
=-(1/2)\omega _{ij}dx^i\wedge dx^j${\it .}

\vspace{0.5cm}\noindent
{\bf Theorem 2.2. }{\it For any observable }$a\in C^\infty (M)[[\hbar ]]$%
{\it \ there is a unique element }$\widetilde{a}\in {\cal A}_D${\it \ such
that }$\sigma (\widetilde{a})=a${\it . Therefore, }$\sigma ${\it \
establishes an isomorphism between }${\cal A}_D${\it \ and }$C^\infty
(M)[[\hbar ]]${\it .}

\vspace{0.5cm}\noindent
{\bf Corollary 2.1}. {\it The pull-back of }$\circ ${\it -product via }$%
\sigma ^{-1}${\it \ induce an associative }$*${\it -product on the space of
physical observables }$C^\infty (M)[[\hbar ]],${\it \ namely} 
\begin{equation}  \label{b16}
a*b=\sigma (\sigma ^{-1}(a)\circ \sigma ^{-1}(b))
\end{equation}
Besides the fact of existence these theorems provide an effective procedure
for the construction of the lifting map $\sigma ^{-1}$ by iterating a pair
of coupled equations

\begin{equation}  \label{iter}
\begin{array}{c}
r=\delta ^{-1}(R+\nabla r+\frac 1{i\hbar }r\circ r), \\ 
\\ 
\sigma ^{-1}(a)=a+\delta ^{-1}(\nabla \sigma ^{-1}(a)+\frac 1{i\hbar
}[r,\sigma ^{-1}(a)])
\end{array}
\end{equation}
Since the operator $\nabla $ preserves the filtration and $\delta ^{-1}$
raises it by $1$, the iteration of the system (\ref{iter}) converges in the
topology (\ref{b3}) and define the unique solution for $\sigma ^{-1}(a)$.
Thus the $*$-product of two functions can be computed with any prescribed
accuracy in $\hbar $.

It should be noted that for the anti-Hermitian $\Lambda $'s (the case of
K\"ahler manifolds) the introduced multiplication possesses the following
property of reality: 
\begin{equation}
\overline{a*b}=\overline{b}*\overline{a},\qquad \forall a,b\in C^\infty
(M)[[\hbar ]]  \label{b17}
\end{equation}
In particular, the formal functions whose coefficients at each power of $%
\hbar $ are real-valued smooth functions form a closed Lie subalgebra with
respect to the $*$-commutator multiplied by $i$. In the standard
quantum-mechanical interpretation, this algebra is usually referred to as an
algebra of {\it physical} {\it observables} corresponding to the algebra of
self-adjoint operators. The formula (\ref{b17}) trivially follows from the
analogous relation for $\circ -$product, 
\begin{equation}
\overline{a\circ b}=(-1)^{mn}\overline{b}\circ \overline{a},\qquad \forall
a\in {\cal A}_{\bullet ,m},~b\in {\cal A}_{\bullet ,n},
\end{equation}
and from the structure of the equations (\ref{b16}), (\ref{iter}).

\vspace{0.5cm}\noindent
{\bf Remark. }As is seen, the rank condition imposed on $\Lambda $ by
Definition 1.1 is certainly inessential for the construction of associative $%
*$-product obeying (\ref{bb1}). The only fact we have used here is that of
existence of a torsion-free connection preserving $\Lambda $. So, the
construction described above works in a more general situation including the
case of degenerate $g$, when $g=0$ we get the Fedosov quantization. It would
be interesting to formulate the necessary and sufficient conditions for a
torsion-free connection preserving a given form $\Lambda $ to exist and to
describe all such connections.

In conclusion let us introduce one more important ingredient of deformation
quantization - a trace functional on the algebra of quantum observables.

\vspace{0.5cm}\noindent
Let $C_0^\infty (M)[[\hbar ]]\subset C^\infty (M)[[\hbar ]]$ be an ideal
consisting of compactly supported quantum observables. Recall, that linear
functional on $C_0^\infty (M)[[\hbar ]]$ with values in formal constants $%
{\Bbb C}[[\hbar ]]$ is called a {\it trace} if it vanishes on commutators%
\footnote{%
Our definition defers from the standard one by the lack of inessential
normalization multiplier $1/(2\pi \hbar )^n.$}, that is 
\begin{equation}
tr(a)=\sum\limits_{k=0}^\infty \hbar ^kc_k,\qquad c_k\in {\Bbb C}
\label{trace}
\end{equation}
and 
\begin{equation}
tr(a*b)=tr(b*a)  \label{trc}
\end{equation}
Let $d\mu =d\mu _0+\hbar d\mu _1+\ldots $ be a formal smooth density on $M$,
then integration by $d\mu $ delivers a continuous functional of the form $%
C_0^\infty (M)[[\hbar ]]\ni a\rightarrow \int_Ma\cdot d\mu $. A formal
density is called a {\it trace density} for $*$-product if the functional $%
\int_Ma\cdot d\mu $ is a trace. The work of Nest and Tsygan \cite{NT}
suggests that any continuous trace functional is defined by a trace density.
For the deformation quantization on FW-manifolds a more profound statement
is true.

\vspace{0.5cm}\noindent
{\bf Theorem 2.3. }{\it \ Up to an overall constant factor there exist a
unique trace density associated to the algebra of quantum observables }$%
C^\infty (M)[[\hbar ]]${\it . It has the form }

\begin{equation}
d\mu =(1+\hbar t_1(x)+\hbar ^2t_2(x)+\ldots )\cdot \Omega  \label{trd}
\end{equation}
{\it where }$\Omega =\omega ^n/n!=\sqrt{|\det g|}dx^1\wedge \ldots \wedge
dx^{2n}${\it \ is symplectic (Riemannian) volume form on }$M${\it \ and
coefficients }$t_i(x)${\it \ are polynomials in curvature tensor }$R_{ijkl}$%
{\it \ and its covariant derivatives.}

\vspace{0.5cm}\noindent
{\bf Proof. }This theorem is the analogue of Theorem 5.6.6 in \cite
{FedosovBook} and it can be proved following the same idea based on a
localization principle. Besides, in the next section we will present the
explicit form for an operator establishing a local isomorphism between the
deformation quantization on FW-manifolds and that on the Fedosov
quantization on the corresponding symplectic manifold $(M,\omega )$. Using
this local isomorphism one may derive the trace density (\ref{trd}) from
Fedosov's trace density for symplectic manifold $(M,\omega )$.

\vspace{0.5cm}\noindent
It is pertinent to note that for a homogeneous FW-manifold the corresponding
symmetry group, acting on $M$ transitively by symplectomorphisms, defines
the symmetry group of deformation quantization. In this case one may see
that $d\mu =\Omega $.

\section{The question of equivalence.}

The rich geometry of the FW-manifolds offers at least two different schemes
for their quantization: the Fedosov quantization, which exploits only the
skew-symmetric part of the form $\Lambda $, and the deformation quantization
involving entire form $\Lambda $ via ansatz (\ref{bb1}). For the reasons
mentioned in the Introduction we refer to these quantizations as those of
Weyl and Wick type, respectively. The natural question aroused is that of
whether we have essentially different quantizations or an equivalence
transform may be found to establish a global isomorphism between both
algebras of quantum observables. Below we formulate the necessary and
sufficient conditions for such an isomorphism to exist. As in the general
case, the obstruction for equivalence of two star-products lies in the
second De Rahm cohomology of symplectic manifold and we identify a certain
2-form as its representative.

In order to distinguish Wick-type star product from the Weyl one, all the
constructions related to the former product will be attributed by the
additional symbol $g$ (pointing on non-zero symmetric part $g$ in $\Lambda $%
). In particular, through this section the fibrewise multiplication (\ref{b2}%
) will be denoted by $\circ _g$, while $\circ $ will be reserved for the
Fedosov $\circ -$product \cite{Fedosov1} resulting from (\ref{b2}) if put $%
g=0$.

\vspace{0.5cm}\noindent
We start by noting that fibrewise $\circ $ and $\circ _g$ products are
equivalent in the following sense:

\vspace{0.5cm}\noindent
{\bf Lemma 3.1.}{\it \ For }$a,b\in {\cal A}${\it \ we have } 
\begin{equation}
a\circ _gb=G^{-1}(G\,a\circ G\,b)  \label{equq}
\end{equation}
{\it where the formal operator }$G${\it \ and its formal inverse are given
by } 
\begin{equation}
G=\exp (-\frac{i\hbar }4g^{ij}\frac \partial {\partial y^i}\frac \partial
{\partial y^j}),\qquad G^{-1}=\exp (\frac{i\hbar }4g^{ij}\frac \partial
{\partial y^i}\frac \partial {\partial y^j})  \label{G}
\end{equation}
{\it In other words, the map }$G:{\cal A}\rightarrow {\cal A}${\it , being
considered as an automorphism of a linear space, establishes the isomorphism
of algebras }$({\cal A},\circ )${\it \ and }$({\cal A}${\it ,}$\circ _g).$

\vspace{0.5cm}\noindent
The proof is obvious from the direct substitution (\ref{G}) to (\ref{equq}).

\vspace{0.5cm}\noindent
The operator $G$ satisfies following simple properties: 
\begin{equation}
\nabla G=G\nabla ,\qquad \delta G=G\delta  \label{prop}
\end{equation}
The automorphism $G$ defines a new Abelian connection $\tilde D=GD_gG^{-1}$
which in virtue of relations (\ref{prop}) can be written as 
\begin{equation}
\tilde D=\nabla -\delta +\frac 1{i\hbar }[\widetilde{r},\cdot ],\qquad 
\widetilde{r}=Gr_g,  \label{newD}
\end{equation}
and the bracket stands for $\circ $-commutator. The elements $\widetilde{r}$
and $r$ fulfill the equations 
\begin{eqnarray}
&&\ \ GR+\nabla \widetilde{r}-\delta \widetilde{r}+\frac 1{i\hbar }%
\widetilde{r}\circ \widetilde{r}  \label{FedeqS} \\
&&\ \ R+\nabla r-\delta r+\frac 1{i\hbar }r\circ r  \label{Fedeq}
\end{eqnarray}
Thus we have two star-products $*$ and $\widetilde{*}$ corresponding to the
pair of Abelian connections $D$ and $\widetilde{D}$. Since $D\neq \widetilde{%
D}$, in general, the action of the fibrewise isomorphism $G$ establishing
the equivalence between $\circ $ and $\circ _g$-products (and hence between
star products $\widetilde{*}$ and $*_g$) did not automatically followed by
the equality $*=\widetilde{*}$ . Indeed, evaluating lowest orders in $\hbar $
we get: 
\begin{eqnarray}
a*b &=&ab+\frac{i\hbar }2\omega ^{ij}\nabla _ia\nabla _jb-\frac{\hbar ^2}%
2\omega ^{ik}\omega ^{jl}\nabla _i\nabla _ja\nabla _k\nabla _lb+{\cal O}%
(\hbar ^3)  \notag \\
&&  \label{sts} \\
a\widetilde{*}b &=&ab+\frac{i\hbar }2\widetilde{\omega }^{ij}\nabla
_ia\nabla _jb-\frac{\hbar ^2}2\omega ^{ik}\omega ^{jl}\nabla _i\nabla
_ja\nabla _k\nabla _lb+{\cal O}(\hbar ^3)  \notag
\end{eqnarray}
where 
\begin{equation}
\widetilde{\omega }^{ij}=\omega ^{ij}+\hbar \omega _1^{ij},  \label{def}
\end{equation}
and 
\begin{equation}
\omega _1^{ij}=\omega ^{ik}\Omega _{kl}\omega ^{lj},\quad \Omega =\frac
i\hbar (GR-R)=\frac 18R_{ijkl}g^{ij}dx^k\wedge dx^l  \label{OmR}
\end{equation}
The 2-form $\Omega $ is closed in virtue of the Bianchi identity for the
curvature tensor. In fact, rels. (\ref{sts}), (\ref{def}) say that the
second star product $\widetilde{*}$ is a so-called {\it 1-differentiable
deformation }of the first one $*$ \cite{Stern}. {\it This deformation is
known to be trivial iff the 2-form }$\Omega ${\it \ is exact }\cite{Stern}, 
\cite{Bonneau}. Now supposing that $\Omega =d\psi $ let us try to establish
an equivalence between $*$ and $\widetilde{*}$ by means of a fibrewise
conjugation automorphism

\begin{equation}
a\rightarrow U\circ a\circ U^{-1},  \label{iU}
\end{equation}
where $U$ is an invertible element of ${\cal A}_{\bullet ,0}$. The element $%
U $ is so chosen that (\ref{iU}) turns $D$ to $\tilde D$. In other words, we
subject $U$ to the condition 
\begin{equation*}
D(U\circ a\circ U^{-1})=U\circ (\tilde Da)\circ U^{-1},\quad \forall a\in 
{\cal A}
\end{equation*}
or, equivalently, 
\begin{equation}
\lbrack U^{-1}\circ DU,a]=\frac 1{i\hbar }[\Delta r,\,a],  \label{Urav}
\end{equation}
where $\Delta r=\widetilde{r}-r$. The last condition means that 
\begin{equation}
U^{-1}\circ DU=\frac 1{i\hbar }\Delta r+\frac 1{i\hbar }\psi ,  \label{urav}
\end{equation}
where $\psi $ is a globally defined 1-form on $M$. The compatibility
condition for equation (\ref{urav}) resulting from the identity $D^2=0$
requires that 
\begin{equation}
DU^{-1}\circ DU=DU^{-1}\circ U\circ U^{-1}\circ DU=-\frac 1{(i\hbar
)^2}\Delta r\circ \Delta r=\frac 1{i\hbar }D\Delta r+\frac 1{i\hbar }d\psi .
\end{equation}
That is 
\begin{equation}
D\Delta r+\frac 1{i\hbar }\Delta r\circ \Delta r+d\psi =0  \label{sovm}
\end{equation}
The analogous relation is obtained if we subtract (\ref{FedeqS}) from (\ref
{Fedeq}) 
\begin{equation}
D\Delta r+\frac 1{i\hbar }\Delta r\circ \Delta r+\Omega =0  \label{compat}
\end{equation}
Comparing (\ref{sovm}) with (\ref{compat}) we conclude that the
compatibility condition holds provided $\Omega $ is exact. Now rewrite (\ref
{urav}) in the form 
\begin{equation}
\delta U=\nabla U+\frac 1{i\hbar }[r,U]-\frac 1{i\hbar }U\circ (\Delta
r+\psi )  \label{urav2}
\end{equation}
and apply the operator $\delta ^{-1}$ to both sides of the equation. Using
the Hodge-De Rham decomposition (\ref{b10}) and taking $\sigma (U)=1,$ we
get 
\begin{equation}
U=1+\delta ^{-1}(\nabla U+\frac 1{i\hbar }[r,U]-\frac 1{i\hbar }U\circ
(\Delta r+\psi ))  \label{proce}
\end{equation}
In \cite[Theorem4.3]{Fedosov1} it was proved that iterations of the last
equation yield a unique solution for (\ref{urav2}) provided the
compatibility condition (\ref{sovm}) is fulfilled$.$ Starting from 1, this
solution defines an invertible element of ${\cal A}_{\bullet ,0}$. Then the
equivalence transform $T:(C^\infty (M),*)\rightarrow (C^\infty (M),*_g)$ we
are looking for is defined as the sequence of maps 
\begin{equation}
Ta(x)=(U\circ G(\sigma _g^{-1}(a))\circ U^{-1})|_{y=0},  \label{on}
\end{equation}
so that 
\begin{equation*}
a*_gb=T^{-1}(Ta*Tb)
\end{equation*}
The main results of this section can be summarized as follows:

\vspace{0.5cm}\noindent
{\bf Theorem 3.1. }{\it The obstruction to equivalence between Weyl and Wick
type deformation quantizations lies in the second De Rham cohomology }$%
H^2(M) ${\it . The quantizations are equivalent iff the 2-form }$%
R_{ijkl}g^{ij}dx^k\wedge dx^l${\it \ is exact}.

\vspace{0.5cm}\noindent
{\bf Remark.} For the anti-Hermitian $\Lambda $ the 2-form $\Omega $ is
nothing but the Ricci form of the K\"ahler manifold $(M,\Lambda )$. In this
case the cohomology class of $\Omega $, being proportional to the first
Chern class $c_1(M)$, is known to depend only on the complex structure of
the manifold \cite{Chern}. Since, for example, $c_1({\Bbb {C}P^n)\neq 0}$
and for any K\"ahler manifold $M$ a topological equivalence $M\sim {\Bbb {C}%
P^n}$ implies a bi-holomorphic one \cite{Ko-Hi}, the Weyl and Wick
quantizations on ${\Bbb {C}P^n}$ are not equivalent for any $\Lambda $.

\section{Superextension.}

In this section we show that the Wick-type deformation quantization of
FW-manifolds described above admits a surprisingly simple generalization to
a certain class of supersymplectic supermanifolds. According to the
Rothstein theory \cite{R} any supersymplectic supermanifold can be
completely specified (up to isomorphism) by the set $(M,\omega ,{\cal E}%
,g,\nabla ^g)$, where $(M,\omega )$ is the symplectic manifold, and ${\cal E}
$ is a vector bundle over $M$ with metric $g$ and $g$-compatible connection $%
\nabla ^g$. When ${\cal E}=TM$ and $M$ is an FW-manifold all Rothstein's
data are already contained in the FW-structure $\Lambda $, and thus, one may
speak about canonical superextension for the FW-manifolds.

As is well known, the geometry of a manifold can be recovered from the
commutative algebra of functions on it. A supermanifold is defined by
extending of such an algebra to a supercommutative superalgebra denoted
below by ${\cal C}$. Geometrically, the elements of ${\cal C}$ can be viewed
as the sections of a Grassmann bundle associated to some vector bundle $%
{\cal E}\rightarrow M$ over a given manifold $M$, that is ${\cal C}=\Gamma
(\Lambda {\cal E})$, and the supercommutative multiplication is given by the
pointwise wedge product, 
\begin{equation}
a\wedge b=(-1)^{d_1d_2}b\wedge a
\end{equation}
Here $a,b\in \Gamma (\Lambda {\cal E}),$ $a$ of degree $d_1$ and $b$ of
degree $d_2$. The elements of superalgebra ${\cal C}$ will be called
superfunctions depending on the commuting coordinates $\{x^i\}$ (i.e. local
coordinates on the base $(M,\Lambda )$) and ``anticommuting coordinates'' $%
\{\theta ^i\}$, so that the general element of superalgebra looks like 
\begin{equation}
{\cal C}\ni a=a(x,\theta )=\sum\limits_{k=1}^{2n}a(x)_{i_1\ldots i_k}\theta
^{i_1}\ldots \theta ^{i_k},
\end{equation}
where $a_k=a(x)_{i_1\ldots i_k}dx^{i_1}\wedge \ldots \wedge dx^{i_k}$
transforms as a $k$-form on $M.$ As usual the symbols $\overset{\rightarrow 
}{\partial }/\partial \theta ^i$ and $\overset{\leftarrow }{\partial }%
/\partial \theta ^i$ will denote the left and right partial derivatives in $%
\theta $'s, which definition is as follows: 
\begin{equation}
\frac{\overset{\rightarrow }{\partial }a}{\partial \theta ^j}%
=\sum\limits_{k=1}^{2n}ka(x)_{ji_2\ldots i_k}\theta ^{i_2}\ldots \theta
^{i_k},\quad \frac{\overset{\leftarrow }{\partial }a}{\partial \theta ^j}%
=\sum\limits_{k=1}^{2n}ka(x)_{i_1\ldots i_{k-1}j}\theta ^{i_1}\ldots \theta
^{i_{k-1}}
\end{equation}
In terms of supercoordinates $(x^i,\theta ^j)$ the Rothstein supersymplectic
form on the superextended FW-manifold $(M,\Lambda )$ is given by 
\begin{equation*}
\Omega =\omega _{ij}dx^idx^j+\frac 12g_{ij}R_{mkl}^j\theta ^m\theta
^idx^kdx^l+g_{ij}D\theta ^iD\theta ^j
\end{equation*}
\begin{equation*}
D\theta ^i\equiv d\theta ^i-\Gamma _{jk}^i\theta ^kdx^j
\end{equation*}
Hereafter we use the following conventions: 
\begin{equation}
dx^idx^j=-dx^jdx^i,\quad \theta ^i\theta ^j=-\theta ^j\theta ^i,\quad
dx^i\theta ^j=-\theta ^jdx^i  \label{con}
\end{equation}
\begin{equation*}
d\theta ^id\theta ^j=d\theta ^jd\theta ^i,\quad \theta ^id\theta ^j=\theta
^jd\theta ^i
\end{equation*}
The deformation quantization of general supersymplectic supermanifolds
defined by Rothstein's data was first performed by Bordemann \cite{SBor}.
The remarkable feature of this construction is that the deformation
quantization for the algebra of superfunctions is performed first and the
super Poisson bracket arises here {\it a posteriori }as{\it \ }$\hbar $%
-linear term of the supercommutator. Below we show how this quantization can
be extended to generate a super Wick symbols associated to the super
FW-manifold. In view of natural bijection between superfunctions and
inhomogeneous differential forms, 
\begin{equation*}
a(x)_{i_1\ldots i_k}\theta ^{i_1}\ldots \theta ^{i_k}\Leftrightarrow
a(x)_{i_1\ldots i_k}dx^{i_1}\wedge \ldots \wedge dx^{i_k},
\end{equation*}
this construction may also be thought of as a deformation of the exterior
algebra of differential forms on $M\footnote{%
It is pertinent to note that in the flat case the deformation of exterior
form calculus was previously considered in \cite{GR, Re}.}.$

To begin with, we define a superextension of the formal symbol algebra $%
{\cal A}$\thinspace introduced in the Definition 2.1\thinspace .

\vspace{0.5cm} \noindent
{\bf Definition 4.1. }{\it The bi-graded superalgebra }${\cal SA}=\oplus
_{m,n=0}^\infty ({\cal SA})_{m,n}${\it \ with unit over }${\Bbb {C}}$ is{\it %
\ a space of formal series, } 
\begin{equation}
a(x,\theta ,y,dx,\hbar )=\sum_{2k+p\ge 0}\hbar ^ka_{k\,i_1\ldots
i_pj_1\ldots j_{q^{\prime }}l_1\ldots l_{q^{\prime \prime
}}}(x)y^{i_1}\ldots y^{i_p}\theta ^{j_1}\ldots \theta ^{j_{q^{\prime
}}}dx^{l_1}\ldots dx^{l_{q^{\prime \prime }}},  \label{sect}
\end{equation}
{\it multiplied with the help of associative }$\circ ${\it -product of the
form} 
\begin{equation}
a\circ b=
\end{equation}
\begin{equation*}
=exp\frac{i\hbar }2\Lambda ^{ij}(x)\left( \frac \partial {\partial y^i}\frac
\partial {\partial z^j}+\frac{\overset{\leftarrow }{\partial }}{\partial
\theta ^i}\frac{\overset{\rightarrow }{\partial }}{\partial \chi ^j}\right)
a(x,y,\theta ,dx,\hbar )b(x,z,\chi ,dx,\hbar )|_{y=z,\theta =\chi }.
\end{equation*}
{\it A general term of the series (\ref{sect}) is assigned by the bi-degree }%
$(2k+p+q^{\prime },q^{\prime \prime })${\it .}

\vspace{0.5cm} \noindent
The associativity of this $\circ $-product is well known (see e.g. 
\cite[Prposition 2.1.]{SBor}) and may be checked by straightforward
computation. As before the expansion coefficients $a_{k\,i_1\ldots
i_pj_1\ldots j_{q^{\prime }}l_1\ldots l_{q^{\prime \prime }}}(x)$ are
considered to be components of covariant tensors on $M$ symmetric in $%
i_1,\ldots ,i_p$ and antisymmetric with respect to $j_1\,,\ldots
\,,j_{q^{\prime }}$ and $l_1\,,\ldots \,,l_{q^{\prime \prime }}$\thinspace .
In accordance with (\ref{con}) one may regard $dx^i$ and $\theta ^j$ as the
set of $2n$ anti-commuting variables, that turns ${\cal SA}$ to ${\Bbb {Z}_2}
$-graded algebra with respect to the Grassmann parity $q=q^{\prime
}+q^{\prime \prime }$ (\ref{sect}). The supercommutator of two homogeneous
elements $a,b\in {\cal SA}$ with the parities $q_1$and $q_2$ is defined as 
\begin{equation}
\lbrack a,b]=a\circ b-(-1)^{q_1q_2}b\circ a.
\end{equation}
The nilpotent operator $\delta $ defined by the equation (\ref{b4}) is still
the inner derivation of the superalgebra ${\cal SA}$ and $\delta ^{-1}$ (\ref
{b9}) is the partial homothopy operator for $\delta $ in the sense of
``Hodge-De Rham'' decomposition (\ref{b10}), where now $\sigma
(a)=a(x,\theta ,0,\hbar ).$ Thus the nontrivial cohomology of $\delta $
coincides with the space of superobservables ${\cal C}[[\hbar ]].$
Obviously, ${\cal SA}$ contains ${\cal A}$ as a subalgebra. The
superextension of covariant derivative (\ref{b11}) from ${\cal A}$ to ${\cal %
SA}$ is defined as follows:

\begin{equation}
\nabla :({\cal SA})_{n,m}\rightarrow ({\cal SA})_{n,m+1},  \label{na}
\end{equation}
\begin{equation*}
\nabla =dx^i\left( \frac \partial {\partial x^i}-y^j\Gamma _{ij}^k(x)\frac
\partial {\partial y^k}-\theta ^j\Gamma _{ij}^k(x)\frac{\overset{\rightarrow 
}{\partial }}{\partial \theta ^k}\right) ,
\end{equation*}
$\Gamma _{ij}^k$ are Christoffel symbols of the connection associated to $%
\Lambda $. Then 
\begin{equation}
\nabla (a\circ b)=\nabla a\circ b+(-1)^qa\circ \nabla b,  \label{deriv}
\end{equation}
with $q$ being the Grassmann parity of $a$. It is easy to check that $\nabla 
$ anti-commutes with $\delta $ and that its curvature is given by

\begin{equation}
\nabla ^2a=\nabla (\nabla a)=\frac 1{i\hbar }[R\,,\,a],  \label{n^2}
\end{equation}
\begin{equation}
R=\frac 14R_{ijkl}y^iy^jdx^kdx^l+\frac 14{\cal R}_{ijkl}\theta ^i\theta
^jdx^kdx^l,  \label{sr}
\end{equation}
where we have used the notations $R_{ijkl}=\omega _{im}R_{~jkl}^m$ and $%
{\cal R}_{ijkl}=g_{im}R_{~jkl}^m$. By analogy with nonsuper case, one can
combine $\nabla $ with an inner derivative to get the Abelian connection of
the form

\begin{equation}
D=\nabla -\delta +\frac 1{i\hbar }[r,\cdot ]=\nabla +\frac 1{i\hbar }[\omega
_{ij}y^idx^j+r,\cdot ],\quad r=r_i(x,\theta \,,y,\hbar )dx^i,  \label{asd}
\end{equation}
\begin{equation*}
D^2a=\frac 1{i\hbar }[\Omega ,a]=0,\quad \forall a\in {\cal SA},\quad \Omega
=-\frac 12\omega _{ij}dx^idx^j,
\end{equation*}
Denote ${\cal SA}_D=\ker D\cap {\cal SA}_{\bullet ,0}$. The next assertion
is the super counterpart of that stated in Theorems 2.1, 2.2. and Corollary
2.1.

\vspace{0.5cm} \noindent
{\bf Theorem 4.1. }{\it With the above definitions and notations we have:}

{\it i) there is a unique Abelian connection }$D${\it \ (\ref{asd}) for
which } 
\begin{equation*}
\delta ^{-1}r=0,\quad r_i(x,\theta \,,0,\hbar )=0
\end{equation*}

{\it and }$r${\it \ consists of monomials whose first degree is no less then
3;}

{\it ii) }${\cal SA}_D${\it \ is a subalgebra of }${\cal SA}${\it \ and the
map }$\sigma ${\it \ being restricted to }${\cal SA}_D$

{\it defines a linear bijection onto }${\cal C}[[\hbar ]]${\it ;}

{\it iii) the formula} 
\begin{equation*}
a*b=\sigma (\sigma ^{-1}(a)\circ \sigma ^{-1}(b))
\end{equation*}

{\it defines the associative multiplication on }${\cal C}[[\hbar ]]$.

\vspace{0.5cm} \noindent
{\bf Proof }can be directly read off from \cite[Theorems 2.1,2.2]{SBor}.

\vspace{0.5cm} \noindent
As in nonsuper case the explicit construction for the lifting map $\sigma
^{-1}$ results from iterations of two coupled equations (\ref{iter}) with
respect to the first degree, with the only difference that $R$ is now
determined by (\ref{sr}) and $a\in {\cal C}[[\hbar ]].$ The first equation
of (\ref{iter}) implies that the element $r$ is an odd one and, as the
result, the lift $\sigma ^{-1}:{\cal C}[[\hbar ]]\rightarrow {\cal SA}_D$
preserves the Grassmann parity.

Using two transverse polarizations on $(M,\Lambda )$, associated to the left
and right kernel distributions of $\Lambda $, one may introduce yet another
graded structure on ${\cal SA}$. In order to present this gradation in
explicit form, let us introduce a frame of (complex) 1-forms $\{e^\alpha
=e_i^\alpha dx^i,$ $f^\beta =f_i^\beta dx^i\}$, $\alpha ,\beta =1,\ldots
,\dim M/2,$ spanned the complexified cotangent bundle over a contractible
coordinate chart $U$. The 1-forms can be chosen to satisfy 
\begin{equation}
Ie^\alpha =e^\alpha ,\quad If^\beta =-f^\beta ,  \label{fr}
\end{equation}
$I$ being the integrable involution structure defined by (\ref{a3}). Then
one can introduce the new (polarized) basis in the space of Grassmann
generating elements $\theta ^i$: 
\begin{equation}
\vartheta ^\alpha =e_i^\alpha \theta ^i,\quad \overline{\vartheta }^\beta
=f_j^\beta \theta ^j  \label{tpm}
\end{equation}
With respect to this basis the superalgebra is decomposed onto the direct
sum of its subspaces 
\begin{equation}
{\cal SA}=\bigoplus\limits_{k=-\infty }^\infty {\cal SA}^{(k)}  \label{dec}
\end{equation}
\begin{equation*}
{\cal SA}^{(k)}\ni a=\sum\limits_{m-n=k}a_{\alpha _1\ldots \alpha _m\beta
_1\ldots \beta _n}(x,y,\hbar ,dx)\vartheta ^{\alpha _1}\ldots \vartheta
^{\alpha _m}\overline{\vartheta }^{\beta _1}\ldots \overline{\vartheta }%
^{\beta _n}
\end{equation*}
The easiest way to see that the decomposition (\ref{dec}) does really define
a ${\Bbb {Z}}$ -gradation with respect to $\circ $-product is to introduce
an inner derivative of the form 
\begin{equation}
\widehat{N}a=\frac 1{i\hbar }[N,a],\quad N=-\frac 12\omega _{ij}\theta
^i\theta ^j
\end{equation}
The main properties of the derivative $\widehat{N}$ are collected in the
following

\vspace{0.5cm} \noindent
{\bf Proposition 4.1.}{\it \ Let all notations be as above, then}

{\it i)} $\widehat{N}a=\theta ^iI_i^k\overset{\rightarrow }{\partial }%
a/\partial \theta ^k;$

{\it ii)} $\widehat{N}a=na,\quad \forall a\in {\cal SA}^{(n)},${\it \ in
particular }$\widehat{N}\vartheta ^\alpha =\vartheta ^\alpha ,\widehat{N}%
\overline{\vartheta }^\alpha =-\overline{\vartheta }^\alpha $;

{\it iii)} $\widehat{N}D=D\widehat{N}\Rightarrow r\in {\cal SA}^{(0)}$;

iv) $\sigma ^{-1}\widehat{N}=\widehat{N}\sigma ^{-1}$.

\vspace{0.5cm} \noindent
The second assertion presents the intrinsic definition of the ${\Bbb {Z}-}$%
gradation (\ref{dec}) without resorting to the special local frame (\ref{fr}%
).

\vspace{0.5cm} \noindent
{\bf Proof. }$i)$ Straightforward computation leads to

\begin{equation}
\widehat{N}a=\frac 1{i\hbar }[N,a]=\theta ^iI_i^k\frac{\overset{\rightarrow 
}{\partial }a}{\partial \theta ^k}-\frac{i\hbar }8\omega _{ij}(\Lambda
^{jk}\Lambda ^{il}-\Lambda ^{kj}\Lambda ^{li})\frac{\overset{\rightarrow }{%
\partial }}{\partial \theta ^l}\frac{\overset{\rightarrow }{\partial }}{%
\partial \theta ^k}a.  \label{n}
\end{equation}
The second term in (\ref{n}) vanishes due to the identities $\omega
_{ij}\Lambda ^{jk}\Lambda ^{il}=\omega _{ij}\Lambda ^{kj}\Lambda ^{li}=0$.

$ii)$ This immediately follows from $i)$ and definition (\ref{tpm}).

$iii)$ We have to show that $DN=0$ or, more explicitly, 
\begin{equation}
\nabla N-\delta N+\frac 1{i\hbar }[r,N]=0
\end{equation}
The first two terms vanishes since $\nabla $ preserves the symplectic
structure $\omega $ and $N$ does not depend on $y$'s. So, it remains to
prove that $N$ commutes with $r$ or, in other words, that $r\in {\cal SA}%
^{(0)}$. The element $r$ is determined from the equation (\ref{iter}), where 
$R$ is given by (\ref{sr}). Expending $r$ with respect to the first degree, $%
r=\sum_{i=3}^\infty r_i,$ $r_i\in {\cal SA}_{i,1}$, and substituting to the
first equation (\ref{iter}) we get a recursive definition for $r$: 
\begin{equation}
r_3=\delta ^{-1}R=\frac 18R_{ijkl}y^iy^jy^kdx^l+\frac 14{\cal R}%
_{ijkl}\theta ^i\theta ^jy^kdx^l,  \label{r3}
\end{equation}
\begin{equation}
r_{3+k}=\delta ^{-1}\left( \nabla r_{2+k}+\frac i\hbar
\sum\limits_{j=1}^{k-1}r_{j+2}\circ r_{k-j+2}\right) ,\quad k=1,2,\ldots .
\label{rrec}
\end{equation}
Since $\widehat{N}$ commutes with $\nabla $ and $\delta ^{-1}$ and
differentiates $\circ $-product the assertion may be proved by induction on
the degree. The starting point of induction is

\begin{equation}
\widehat{N}r_3=\frac 1{i\hbar }[N,\delta ^{-1}R]=-\frac 12R_{ijkl}y^k\theta
^i\theta ^jdx^l=0.
\end{equation}
The latter equality holds because $R_{ijkl}=R_{jikl}$.

$iv)$ Let $\widetilde{a}=\sigma ^{-1}(a)$. We expand $a$ and $\widetilde{a}$
with respect to the first degree: 
\begin{equation}
a=\sum_{k=0}^\infty a_k,\quad \widetilde{a}=\sum_{k=0}^\infty \widetilde{a}%
_k,\quad a_k=\hbar ^k\alpha _k,\quad \alpha _k\in {\cal C},\quad \widetilde{a%
}_k\in {\cal SA}_{k,0}.  \label{exp}
\end{equation}
Given $r$, then the sequence $\{\widetilde{a}_k\}$ is recursively determined
via $\{a_k\}$ from the equation (\ref{iter}) as follows: 
\begin{equation}
\widetilde{a}_k=a_k+\delta ^{-1}\left( \nabla (\widetilde{a}_{k-1})+\frac
i\hbar \sum\limits_{n=1}^{k-2}[r_{n+2},\widetilde{a}_{k-n-1}]\right) ,\quad
k=1,2,\ldots ,  \label{lift}
\end{equation}
where of course an empty sum is assumed to be zero and $\widetilde{a}_0=a_0.$
Applying $\widehat{N}$ to both sides of (\ref{lift}) and using the fact that 
$\widehat{N}r=0$ we get a chain of equations determining the lift of $%
\widehat{N}a$. But this means{\it \ iv)}.

\vspace{0.5cm} \noindent
Now let us clarify the relationship between the algebra of quantum
observables on FW-manifold $(C^\infty (M)[[\hbar ]],*^{\prime }),$
introduced in Section 3, and its superextension $({\cal C}[[\hbar ]],*)$. In
order to avoid a confusion all the constructions related to the former
algebra will be labelled by prime. First of all, we note that the subspace $%
C^\infty (M)[[\hbar ]]\subset {\cal C}[[\hbar ]]$ is not closed with respect
to $*$-product. The reason is that even though elements $a$ and $b$ do not
depend on $\theta $'s there lifts $\sigma ^{-1}(a)$ and $\sigma ^{-1}(b),$
defined with respect to (\ref{asd}), are in general do and so does their
product{\bf .} Thus, the natural embedding $C^\infty (M)[[\hbar ]]\subset 
{\cal C}[[\hbar ]]$ does not induce a homomorphism of algebras; instead, the
following relation holds true: 
\begin{equation}
a*^{\prime }b=\pi (a*b),\quad \forall a,b\in C^\infty (M)[[\hbar ]],
\end{equation}
where $\pi :{\cal C}[[\hbar ]]\rightarrow C^\infty (M)[[\hbar ]]$ is the
canonical projection defined by the rule: $\pi a(x,y,\hbar ,\theta
)=a(x\,,y\,,\hbar ,0)$. This relation is the particular case of a more
general

~\\[1cm]

\vspace{0.5cm} \noindent
{\bf Proposition 4.2.} {\it For }$\forall ${\it \ }$a\in C^\infty (M)[[\hbar
]]${\it \ and for }$\forall ${\it \ }$b\in {\cal C}[[\hbar ]]${\it \ we have:%
}

\begin{equation}  \label{induce}
a*^{\prime }(\pi b)=\pi (a*b),\quad (\pi b)*^{\prime }a=\pi (b*a)
\end{equation}

\vspace{0.5cm} \noindent
Before proving this proposition consider first the fibrewise analog of
relations (\ref{induce}).

\vspace{0.5cm} \noindent
{\bf Proposition 4.3.}{\it \ Let }$\pi :{\cal SA\rightarrow A}${\it \ be a
canonical projection defined as } \\$\pi a(x,y,\theta ,dx,\hbar
)=a(x,y,0,dx,\hbar )${\it . Consider the algebras }${\cal SA}${\it \ and }$%
{\cal A}${\it \ as left (right) moduli over }${\cal SA}^{(0)}${\it \ and }$%
{\cal A}${\it , respectively. Then }$\pi ${\it \ defines a homomorphism of
the moduli introduced, that is for }$\forall a\in {\cal SA}^{(0)}${\it \ and 
}$\forall b\in {\cal SA}$%
\begin{equation}  \label{idp}
\pi (a\circ b)=(\pi a)\circ (\pi b),\quad \pi (b\circ a)=(\pi b)\circ (\pi a)
\end{equation}

\vspace{0.5cm} \noindent
{\bf Proof. }In view of (\ref{dec}) we have the expansion $%
a=\sum_{n=0}^\infty a_n,$ where 
\begin{equation*}
a_n=a_{\alpha _1\ldots \alpha _n\beta _1\ldots \beta _n}(x,y,\hbar
,dx)\vartheta ^{\alpha _1}\ldots \vartheta ^{\alpha _n}\overline{\vartheta }%
^{\beta _1}\ldots \overline{\vartheta }^{\beta _n}.
\end{equation*}
Now it suffices to prove that for $n>0,$ $\pi (a_n\circ b)=\pi (b\circ a_n)=0
$. But this directly follows from the identities 
\begin{equation}
b\circ \vartheta ^\alpha =b\vartheta ^\alpha ,\quad \overline{\vartheta }%
^\alpha \circ b=\overline{\vartheta }^\alpha b,\quad \forall b\in {\cal SA}.
\end{equation}
To prove Proposition 4.2, we need the following

\vspace{0.5cm} \noindent
{\bf Lemma 4.1.} $\pi \sigma ^{-1}(a)=(\sigma ^{-1})^{\prime }\pi a$, $%
\forall a\in {\cal C}[[\hbar ]].$

\vspace{0.5cm} \noindent
{\bf Proof.} Denote as before $\widetilde{a}=\sigma ^{-1}(a)$. The element $%
\widetilde{a}$, being expanding as in (\ref{exp}), is recursively determined
from the pair of coupled equations (\ref{rrec},\ref{lift}). Note that $\pi
r_3=r_3^{\prime }$ and $\pi $ commutes with $\nabla $ and $\delta ^{-1}.$
Now apply the canonical projection $\pi $ to both sides of equations (\ref
{rrec},\ref{lift}). In doing so, we can use the identities (\ref{idp}) since 
$r\in {\cal SA}^{(0)}$ in view of Proposition 4.1. By induction we have 
\begin{equation}
r_{3+k}^{\prime }=\delta ^{-1}\left( \nabla r_{2+k}^{\prime }+\frac i\hbar
\sum\limits_{j=1}^{k-1}r_{j+2}^{\prime }\circ r_{k-j+2}^{\prime }\right) ,
\end{equation}
\begin{equation}
\pi \widetilde{a}_n=\pi a_n+\delta ^{-1}\left( \nabla (\pi \widetilde{a}%
_{n-1})+\frac i\hbar \sum\limits_{k=1}^{n-2}[r_{k+2}^{\prime },\pi 
\widetilde{a}_{n-1-k}]\right) .
\end{equation}
But this is precisely the recursive definition of $(\sigma ^{-1})^{\prime
}\pi a$. Thus the theorem is proved.

\vspace{0.5cm} \noindent
Now the Proposition 4.2. follows from the sequence of equalities 
\begin{eqnarray*}
\pi (a*b) &=&\pi \sigma (\sigma ^{-1}a\circ \sigma ^{-1}b)=\sigma \pi
(\sigma ^{-1}a\circ \sigma ^{-1}b)=\sigma (\pi \sigma ^{-1}a\circ \pi \sigma
^{-1}b)= \\
&& \\
&=&\sigma ((\sigma ^{-1})^{\prime }\pi a\circ (\sigma ^{-1})^{\prime }\pi
b)=\sigma ((\sigma ^{-1})^{\prime }a\circ (\sigma ^{-1})^{\prime }\pi
b)=a*\pi b
\end{eqnarray*}
and analogous relations for the second equality of (\ref{induce}).

\section{Discussion.}

The deformation quantization, as it was originally formulated, was aimed to
provide the rigor mathematical groundwork for what physicists called
quantization. Nowadays the methods of deformation quantization go far beyond
the quantization problem by itself and constitute an integral and the most
elaborated part of more general concept -- noncommutative geometry \cite{Co}%
. The fresh interest to the subject was provoked by the resent developments
in the M-theory \cite{CDS,SW}, where a certain limit of nonperturbative
string dynamics was recognized as the noncommutative Yang-Mills theory (NYM) 
\cite{CR}. By now, however, only the restricted class of such models has
been intensively studied, namely, the models formulated on linear symplectic
space ${\Bbb R}^n$ or tori ${\Bbb T}^n$ endowed with canonical Poisson
brackets and Euclidean metric. Turning to the more general manifolds (such
as ALE and K3 spaces, Calabi-Yau orbifolds, higher genus Riemann surfaces
etc., which all of physical interest) one has to deal with the non-constant
symplectic and metric structures. It is clear, that the naive
covariantization of the flat NYM Lagrangian destroys the associativity of
star-multiplication and thus it may break the gauge invariance of the
theory. So, the problem is to unify consistently three different ingredients
of the NYM theory: the noncommutativity of the coordinates, governed by some
Poisson structure; the metric properties of the manifold and the tensor
nature of the YM fields. This can be also considered as a part of more
general mathematical problem: given a Poisson manifold, how to define the
associative deformation of the corresponding tensor algebra and fundamental
tensor operations? In the special case of functions, forming the subalgebra
in the full tensor algebra, it is just the problem which is studied by the
deformation quantization.

In the seminal paper \cite{CDS} the Fedosov deformation quantization was
suggested as a possible tool for constructing the NYM on general symplectic
manifolds. So far as we know, only two attempts have been undertaken to
proceed in this direction \cite{HGC}, \cite{AK}, and both are not too
successful. Without going into details we note that the action functional
for NYM proposed in \cite{HGC} is not in fact gauge invariant (except the
trivial flat case) because the Fedosov covariant derivative has not been
properly extended to the tensor observables. In the work\cite{AK} a local
frame of closed 1-forms is introduced to convert the tensor observables into
the scalar functions, which may be then quantized by the Fedosov method.
However, the technical restrictions imposed by the authors actually imply
the existence of global Darboux coordinates on the manifold that makes all
the construction trivial.

We hope that the super FW-manifolds introduced in this work may provide a
geometrical background for constructing NYM theories in nontrivial
symplectic manifolds. Indeed, the natural bijection between differential
forms and superfunctions, 
\begin{equation*}
a(x)_{i_1\ldots i_k}\theta ^{i_1}\ldots \theta ^{i_k}\Leftrightarrow
a(x)_{i_1\ldots i_k}dx^{i_1}\wedge \ldots \wedge dx^{i_k},
\end{equation*}
give rise to the associative deformation of the exterior algebra on the base
FW-manifold. The NYM fields are embedded in this algebra as 1-forms. The
gauge transformations are then associated with the internal automorphisms of
the superalgebra,

\begin{equation*}
\delta _\varepsilon a=[a,\varepsilon ].
\end{equation*}
Note, that subspace of 1-forms is not in general invariant under these
transformations even when the gauge parameter $\varepsilon $ is restricted
to be a function (0-form). Thus a hypothetical NYM theory being consorted in
this way would necessary incorporate the entire multiplet of antisymmetric
tensor fields. In the flat limit, however, the dynamics of NYM field is
expected to decouple from the dynamics of higher-rank forms. The main
advantage of this approach is that the symplectic and metric structures
enter to the formalism on equal footing from the very beginning, that
ensures the general covariance of the theory in question. Finally, the
construction of action functional for NYM theory requires to define the
trace functional on the algebra of superobservables. This is still an open
question and we postpone it to a future work.

The covariant Wick-type symbol construction, given in this paper, may
probably find an application in the problems related to quantization of the
nonlinear field theory models like strings in the curved spaces. The naive
canonical quantization, being based on the Weyl type symbol, is usually
inadequate in this case (as is seen even from the simple case of the bosonic
string in the flat space), while the nontrivial operator formulation is
known only for special examples of nonlinear theories, like WZWN sigma
models. The perturbative quantization of the nonlinear theory might be
sometimes possible by means of the expansion over the linear background
making use of the Wick symbol defined with respect to the linear
approximation (examples of the perturbative construction of the Wick symbols
see in ref. \cite{FLP},\cite{KL}, \cite{KarL} ). However, the expansion over
the linear vacuum can not always be an efficient method, e.g., it would be a
hopeless attempt to construct the string theory in the AdS space taking it
as a weak perturbation to the Minkowskian string vacuum. As one can expect,
the presented Wick symbol formalism may have a potential for constructing an
adequate deformation quantization procedures for the nonlinear field
theoretical models. However, the practice of the general method application
should be developed in separate works for the actual problems of this type.
In particular, the physically relevant models are usually subject to the
phase space constraints, so the symbol construction should be first adapted
to the case of a constrained system. Second, it should be understood in the
nonlinear field theory how one can identify the symmetric tensor which
defines the Wick symbol in the phase space. We expect that in many of
physically relevant cases, the second problem can be reduced to the first
one, because it frequently occurs that a nonlinear theory can be
equivalently represented as a linear one subject to constraints (e.g.: the
AdS string can be thought of as a string on the constrained hypersurface the
in flat space with an additional time dimension).

\section{Acknowledgments}

We are grateful to I.~Gorbunov, M.~Grigoriev, M.~Henneaux, A.~Karabegov,
R.~Marnelius, A.~Nersessian and M.~Vasiliev for valuable discussions. SLL
appreciates partial financial support from the RFBR grant No 99-01-00980,
the work of VAD and AAS is partially supported by the RFBR grant No
00-02-17-956.

\end{document}